\documentclass[aps,twocolumn]{revtex4}
\newcommand{\be}{\begin{eqnarray}}
\newcommand{\ee}{\end{eqnarray}}

\newcommand{\lb}{\label}
\def\>{\rangle}
\def\<{\langle}

\def\tr{\hbox{Tr}}
\def\r{\rho}
\def\be{\begin{eqnarray}}
\def\ee{\end{eqnarray}}
\def\lb{\label}
\begin{document}
\title{Entanglement entropy of black holes and AdS/CFT correspondence}
\author{
Sergey N. Solodukhin
} \affiliation{{\it School of Engineering and Science,
International University Bremen,
 P.O. Box 750561, Bremen  28759, Germany}}

\begin{abstract}
\noindent { A recent proposal by Ryu and Takayanagi for a
holographic
  interpretation of entanglement entropy in conformal field theories dual to
  supergravity on anti-de Sitter (adS) is generalized to include  entanglement entropy
  of black holes living on the boundary of adS. The generalized proposal is
  verified in boundary dimensions $d=2$ and $d=4$ for both the UV divergent and UV
  finite terms. In dimension $d=4$ an expansion of entanglement entropy in terms of
  size $L$ of the subsystem
  outside the black hole is considered.  A new term in the entropy of dual strongly coupled CFT,
  which universally grows  as $L^2\ln L$ and is
  proportional to the value of  the obstruction tensor at the black hole
  horizon, is predicted.

\noindent {PACS: 04.70Dy, 04.60.Kz, 11.25.Hf\ \ \ }}
\end{abstract}
\vskip 2.pc \maketitle

\section{ Introduction.} For more than a decade  entanglement
entropy has remained in the focus of theoretical studies. Its
geometrical nature and the fact that it grows as area rather than
as volume makes entanglement entropy an attractive candidate for
the statistical-mechanical origin of the Bekenstein-Hawking
entropy \cite{Bombelli:1986rw}-\cite{Fursaev:2006ng}. Much of the
past interest was related to understanding the UV divergences of
the entropy that resulted in the demonstration
\cite{Fursaev:1994ea} that these divergences are the same as those
of the quantum action and  are renormalized by the standard
renormalization of the gravitational couplings in the action. This
result guarantees that if the couplings (Newton's constant and
those in front of the quadratic in curvature terms)  are entirely
due to quantum effects (bare gravitational couplings vanish), as
in \cite{Hawking:2000da}, then entanglement entropy is precisely
the Bekenstein-Hawking entropy. Another related approach with
similar behavior of the entropy is the ``brick wall'' model of 't
Hooft \cite{'tHooft:1984re}, \cite{Demers:1995dq}. Entanglement
entropy in string theory was discussed in \cite{Brustein:2005vx}.

Recently, there has been  renewed interest in the study of
entanglement entropy. This interest is two-fold. On one side,
entropy is considered to be a reasonable measure of entanglement
in quantum systems that is needed for the realization of the idea
of quantum computers. This motivates the studies of entanglement
entropy in condensed matter models
\cite{Korepin}-\cite{Riera:2006vj}. On the other side,
entanglement entropy was recently studied in the context of the
holographic AdS/CFT correspondence (for a review of the
correspondence see \cite{Aharony:1999ti}). This study
\cite{Ryu:2006bv}, \cite{Ryu:2006ef} suggests a new geometric (and
completely classical) picture for entanglement entropy of quantum
systems. This offers new possibilities for understanding   the
black hole entropy as entanglement entropy \cite{Emparan:2006ni}.
The proposal of \cite{Ryu:2006bv}, \cite{Ryu:2006ef} relates  the
entanglement entropy of a quantum system living in the boundary of
anti-de Sitter (adS) with the area of the minimal surface in the
adS whose boundary is the surface that divides the subsystems in
the boundary theory. This  proposal works exactly in the d=2 case,
but can only be checked approximately for higher dimensions
because the CFT entanglement entropy is only known in the free
field limit, far from the strongly coupled limit that is expected
to agree with the classical gravity result.
 The success of this proposal for systems in flat space-time
suggests that this may be generalized for space-times containing a
black hole. In that case the black hole would lie in the boundary
of the anti-de Sitter space. One proposal for this generalization
is available \cite{Emparan:2006ni} (see also
\cite{Iwashita:2006zj}) and consists in considering a minimal
surface that bounds the black hole horizon. It is, however, not
clear how this proposal works in two dimensions when the horizon
is just a point. On the other hand, it does not give the expected
behavior of entanglement entropy on the size of the total system.
In this note we suggest another way to generalize the construction
of \cite{Ryu:2006bv}, \cite{Ryu:2006ef} for black holes which
resolves these issues.

\section{ Entanglement entropy of black holes.} Entanglement
entropy is defined for a system divided into two subsystems $A$
and $B$ for a given state determined in general by a density
matrix $\rho(A,B)$. Typically, the total system is considered to
be in a pure state (for instance, in a ground state) described by
wave function $|\Psi>$ then $\rho(A,B)=|\Psi ><\Psi |$. It can
also be in a mixed, for instance, thermal state, described by a
density matrix. If one neglects information about a subsystem, say
$B$, then this situation is described by the reduced density
matrix $\rho_A=\tr_B \rho(A,B)$, where a partial trace over only
subsystem $B$ is taken. Thus, provided we have access only to a
part of the system then  entanglement entropy defined as
$S_A=-\tr_A \rho_A\log \rho_A$ gives us a measure for the lack of
information about the state of the total system. On the other
hand, this entropy is non-vanishing because there are correlations
between  the subsystems. Entanglement entropy, thus, can be also
viewed as some measure for  these correlations. In local theories
the correlations are  short-distance and, hence, the  entropy
 is expected to be determined by the geometry of the boundary dividing the two
subsystems. On the other hand, the entropy should depend on the
geometry (size and shape) of the total system $A\cup B$. In the
limit, when the size of any of the subsystems is large, the
reduced density matrix becomes thermal, so that the entropy grows
as volume,  if the total system  was in a thermal state.

The notion of entanglement entropy is ideally suited for black
holes since the black hole horizon naturally divides  space-time
on two subsystems so that  an observer outside a black hole  does
not have  access to excitations propagating inside the horizon.
Entanglement entropy for black holes, especially the UV divergent
part of the entropy, was well studied in the past (see
\cite{Callan:1994py}-\cite{Solodukhin:1996vx} and references
therein). In two dimensions, for a conformal field theory with
central charge $c$, entanglement entropy can be obtained
explicitly including the UV finite terms. As in the case of a CFT
on interval  \cite{Holzhey:1994we}, \cite{Korepin},
\cite{Calabrese:2004eu}
 the conformal symmetry plays an
important role in providing us with a closed form expression. For
a black hole metric written in the conformally flat form
$g_{\mu\nu}=e^{2\sigma}\delta_{\mu\nu}$,
 entanglement entropy  \cite{RCM}, \cite{Fursaev:1995ef} is given by value of the conformal factor at horizon,
$S={c\over 6}\sigma (x_+)+{c\over 6} \ln (\Lambda/ \epsilon )$,
where $\epsilon$ is a UV regulator and $\Lambda$ is a dimensionful
parameter. Let the black hole geometry be described by a 2d metric
\begin{equation}
ds^2_{bh}=f(x)d\tau^2+{1\over f(x)}dx^2, \label{B}
\end{equation}
where the metric function $f(x)$ has simple zero in $x=x_+$;
$x_+\leq x \leq L$,
 $0\leq \tau
\leq \beta_H$, $\beta_H={4\pi \over f'(x_+)}$. It is easy to see
that (\ref{B}) is conformal to the flat disk of radius $z_0$ ($\ln
z={{2\pi \over \beta_H}\int^x_L {dx \over f(x)}}$):
\begin{eqnarray}
&&ds^2_{bh}=e^{2\sigma }z_0^2(dz^2+z^2d\tilde{\tau}^2 )~~, \\
&&\sigma={1\over 2} \ln f(x)-{2\pi \over \beta_H}\int^x_L {dx\over
f(x)}
 +\ln {\beta_H  \over 2\pi z_0}, \nonumber
\label{4}
\end{eqnarray}
where
 $\tilde{\tau}={2\pi \tau \over \beta_H}$ ($0\leq \tilde{\tau} \leq2\pi$),
 $0 \leq z\leq 1$.
So that entanglement entropy of the 2d black hole takes the form
 \cite{Solodukhin:1996vx}, \cite{Frolov:1996hd}
\begin{equation}
S={c\over 12} \int^L_{x_+}{dx \over f(x)}({4\pi \over
\beta_H}-f')+ {c\over 6}\ln ({\beta_H  f^{1/2}(L)\over \epsilon}),
\label{S}
\end{equation}
where we omit the irrelevant term that is a function of $(\Lambda
, z_0)$ but not of the parameters of the black hole and have
retained dependence on UV regulator $\epsilon$. Notice that the
UV-divergent term here is proportional to $c/6$ rather than $c/3$
as in the case of the field system on interval
\cite{Holzhey:1994we}, \cite{Korepin}. This is because the
interval has two ``boundaries'' where subsystems come into contact
with each other, while in the case of a black hole there is only
one such a boundary (horizon). As was shown in
\cite{Solodukhin:1996vx}, entanglement entropy (\ref{S}) is
identical to the entropy of the thermal atmosphere of quantum
excitations outside the horizon in the 'tHooft's
\cite{'tHooft:1984re} "brick wall" approach.

The black hole resides inside a finite size box  and $L$ is the
coordinate of the boundary of the box. The coordinate invariant
size of the subsystem complimentary to the black hole is $L_{\tt
inv}=\int_{x_+}^Ldx/\sqrt{f(x)}$. Two limiting cases are of
interest.  In the first,  the size of the system $L_{\tt inv}$ is
taken to infinity. Then, assuming that the black hole space-time
is asymptotically flat, we obtain that entanglement entropy
(\ref{S}) approaches the entropy of the thermal gas,  $S_{\tt
th}={c\pi\over 3}L_{\tt inv}T_H$. Another interesting  case is
when  $L_{\tt inv}$ is small. In this case we find the universal
behavior \be S={c\over 6}\left(\ln{L_{\tt inv}\over
\epsilon}-{R(x_+)\over 36}L^2_{\tt
    inv}+O(L^3_{\tt inv})\right) .
\lb{Lsm} \ee The universality of this formula is in the fact that
it does not depend on any characteristics of a  black hole (mass,
temperature) other than the value of the curvature $R(x_+)$ at the
horizon.

In higher dimensions, much is known about the UV-divergent part of
the entropy. The leading UV divergence is proportional to the area
of the horizon while the subleading logarithmically divergent term
is determined by the way the horizon surface
 $\Sigma$ is embedded in
space-time. In four dimensions one has \cite{Solodukhin:1994yz},
\cite{Solodukhin:1994st}, \cite{Fursaev:1994ea} \be
S_{div}={c_0\over 4\epsilon^2}A_\Sigma -8\pi \int_{\Sigma} ( c_{1}
R +{c_{2}\over 2}R_{aa}+ c_{3}R_{abab} ) \ln \epsilon,
\label{Sdiv1} \ee where values of $c_k,\ k=0..3$ depend on the
type of  field, $R_{aa}=R_{\mu\nu}n^\mu_a n^\nu_a$,
$R_{abab}=R_{\mu\nu\alpha\beta}n^\mu_a n^\alpha_a n^\nu_b
n^\beta_b$ and $n^a,\ a=1,2$  is a pair of orthonormal vectors
orthogonal to $\Sigma$. These divergences of the entropy are
related to the UV divergences of the quantum action
\cite{Fursaev:1994ea}, \cite{Fursaev:1995ef}
$$W_{div}=-\int [{c_0\over 16\pi \epsilon^2}  R
- \left(c_{1}R^2+c_{2} R^2_{\mu \nu} +c_{3}
R^2_{\mu\nu\alpha\beta} \right)\ln \epsilon ].
$$

\section{ The proposal.} Our proposal for the holographic
interpretation of  entanglement entropy of a black hole is the
following: Let $d$-dimensional spherically symmetric static black
hole with horizon $\Sigma$
 lie on the regularized boundary (with regularization parameter $\epsilon$)
of asymptotically anti-de Sitter space-time  $adS_{d+1}$ inside a
spherical cavity $\Sigma_L$ of radius $L$. Consider a minimal
d-surface $\Gamma$ whose boundary is the union of $\Sigma$ and
$\Sigma_L$. $\Gamma$ has saddle points where the radial adS
coordinate  has extremum. By spherical symmetry the saddle points
form (d-2)-surface ${\cal C}$ with geometry of a sphere. Consider
the subset $\Gamma_h$ of $\Gamma$ whose boundary is  the union of
$\Sigma$ and ${\cal C}$. According to our proposal, the quantity
\be S={{\tt Area}(\Gamma_h)\over 4 G_N^{d+1}} \lb{proposal} \ee is
equal to entanglement entropy of the black hole in the boundary of
AdS. In particular, it gives the expected dependence of the
entropy on the UV regulator $\epsilon$ and on the size of the
system. Below we verify this proposal for $d=2$ and $d=4$.

\section{Entanglement entropy of 2d black holes from the AdS/CFT
  correspondence.}
In order to check our proposal we need  a solution to the bulk
Einstein equations that describes a black hole in the boundary of
adS. In three dimensions a solution of this type is known
explicitly \cite{Skenderis:1999nb}, \be &&ds^2 = {d\rho^2\over
4\rho^2} +{1 \over \r} \left[f\left(1 + {1\over 16}{f'{}^2- b^2
\over f}\r\right)^2
 d\tau^2 \right.\nonumber \\
&&\left. +{1 \over f}\left( 1 +{1\over 4} f'' \r - {1 \over
16}{f^{' 2}- b^2 \over f}\r\right)^2 dx^2 \right],\label{metric}
\ee where we set the adS radius to unity. At asymptotic infinity
($\r=0$) of  metric (\ref{metric}) one has the 2d black hole
metric (\ref{B}). Note, that (\ref{metric}) is a vacuum solution
of the Einstein equations for any function $f(x)$ in (\ref{B}).
 The regularity requires that constant
$b$ should be related to the Hawking temperature of the
two-dimensional horizon, $b=f'(x_+)$. The geodesic $\Gamma$ lies
in the hypersurface of constant time $\tau$. The induced metric on
the hypersurface $(\rho , x)$  has a constant curvature equal $-2$
for any function $f(x)$ and is, thus, related by coordinate
transformation to metric \be ds_\tau^2={dr^2\over 4r^2}+{1\over
r}dw^2. \lb{ads2} \ee The exact relation between the two
coordinate systems is \be
w={1\over 8b}e^{z(x)}\left({16f(x)-(b^2-f'^2_x)\r\over 16f(x)+(b-f'_x)^2\r}\right)\nonumber \\
r=f(x){e^{2z(x)}}{\r\over (16f(x)+(b-f'_x)^2\r)^2}, \lb{trans} \ee
where $z(x)={b\over 2}\int_L^x {dx\over f(x)}$. The equation for
the geodesic in metric (\ref{ads2}) is $r={1\over C^2}-(w-w_0)^2$.
The geodesic length between two points with radial coordinates
$r_1$ and $r_2$ is $\gamma(1,2)={1\over 2}\ln
\left[{1-\sqrt{1-C^2r}\over 1+\sqrt{1-C^2r}}\right]_{r_1}^{r_2}$.
The saddle point of the geodesic is at $r_m=1/C^2$. The length of
the geodesic $\Gamma_h$ joining point $r_+$ corresponding to
$(x=x_+,\rho=\epsilon)$ and the saddle point is $\gamma
(\Gamma_h)=-{1\over 2}\ln{C^2r_+\over 4}. $
 The constant $C$ is determined from the condition that
geodesic $\Gamma$ joins points $(x=x_+,\rho=\epsilon^2)$ and
$(x=L,\rho=\epsilon^2)$ lying on the regularized (with
regularization parameter $\epsilon$) boundary. One finds that
${C^2 r_+\over 4}=\epsilon^2{b^2\over
f(x_+)}\exp(-b\int^L_{x_+}{dx\over f(x)}) $ in the limit of small
$\epsilon$. Notice, that this is a finite quantity even though
$1/f(x_+)$ itself is divergent. Now, taking into account that
${1\over G_N}={2c\over 3}$, where $c$ is the central charge of
boundary CFT, we find \be
&&S={1\over 4G_N}\gamma(\Gamma_h)  \\
&&={c\over 6}\ln{1\over \epsilon}+ {c\over 12}\left[
\int_{x_+}^L{dx\over f(x)}(b-f')+\ln f(L)-\ln b^2\right]\nonumber
\lb{Sgeom} \ee that coincides with the expression (\ref{S})  for
entanglement entropy of the 2d black hole in conformal field
theory. This confirms our proposal in two dimensions.

\section{ Entanglement entropy of higher dimensional black holes:
UV divergences and finite terms.} In higher dimensions there is no
exact solution similar to (\ref{metric}). However, a solution in
the form of  an asymptotic in $\rho$  expansion is available.  In
the rest of the note we focus on the case of boundary dimension 4.
Then one finds \cite{FG}, \cite{Henningson:1998gx} \be
&&ds^2={d\r^2\over 4\r^2}+{1\over \r}g_{ij}(x,\rho)dx^idx^j \lb{exp1} \\
&&g_{}(x,\rho)=g_{(0)}(x)+g_{(2)}\rho+g_{(4)}\r^2+h_{(4)}\r^2\ln\r
+..,\nonumber \ee where $g_{(0)ij}(x)$ is the boundary metric,
coefficient $g_{(2)ij}=-{1\over 2}(R_{ij}-{1\over 6}Rg_{(0)ij})$
is the local covariant function of boundary metric
\cite{Henningson:1998gx}. Coefficient $g_{(4)}$ is not expressed
as a local function of the boundary metric and is related to the
stress-energy tensor of the boundary CFT \cite{deHaro:2000xn},
which has an essentially nonlocal nature. $h_{(4)}$ is a local
covariant function of the boundary metric and is obtained as a
variation of the integrated conformal anomaly with respect to the
boundary metric \cite{deHaro:2000xn}. Explicit form  was computed
in \cite{deHaro:2000xn} and reads (in curvature conventions with a
different  sign than those used in \cite{deHaro:2000xn}) \be
&&h_{(4)ij}={1\over 8}R_{ikjl}R^{kl}-{1\over 48}\nabla_i\nabla_j
R+{1\over
  16}\nabla^2 R_{ij} \lb{h4} \\
&&-{1\over 24}RR_{ij} +({1\over 96}R^2-{1\over
  32}R_{kl}^2-{1\over 96}\nabla^2 R)g_{(0)ij}. \nonumber
\ee It is a conformal tensor and in mathematics literature  is
known as the ``obstruction" tensor \cite{GH}. We choose the
boundary metric  describing
 a static spherically symmetric black hole to take the form
\be ds^2=f(r)d\tau^2+f^{-1}(r)dr^2+r^2(d\theta^2+\sin^2\theta
d\phi^2) \lb{spher} \ee The minimal surface $\Gamma$ lies in the
hypersurface of the constant $\tau$ of 5-dimensional space-time
(\ref{exp1}).
 The induced metric on the hypersurface takes the form
\be ds^2_{\tau}={d\r^2\over 4\r^2}+{1\over \r}\left[F{dr^2\over
f(r)} +R^2(d\theta^2+\sin^2\theta d\phi^2)\right], \lb{hyper} \ee
where functions $F(r,\rho)=g^{rr}_{(0)}g_{rr}$ and
$R^2(r,\rho)=g_{\theta\theta}$ have $\rho$-expansion due to
(\ref{exp1}). The minimal surface $\Gamma$ can be parameterized by
$(\rho, \theta, \phi)$. Instead of radial coordinate $r$ it is
convenient to introduce coordinate $y=\int dr/\sqrt{f}$ so that
near the horizon one has $r=r_++by^2/4+O(y^4)$. $y$ measures the
invariant distance along the radial direction. By spherical
symmetry, the area to be minimized is \be {\tt Area}(\Gamma)=4\pi
\int {d\r \over \r}R^2 \sqrt{{1\over 4\r^2}+{F\over \r} ({dy\over
d\r})^2}, \lb{*} \ee where $\r_m$ is coordinate of the saddle
point. In the small vicinity of the horizon ($y\ll 1$), we can
neglect dependence of functions $F(y,\r)$ and $R^2(y,\rho)$ on
coordinate $y$. The minimization of the area of the surface gives
the equation \be {FR^2{dy\over d\r}\over \r^2\sqrt{{1\over
4\r^2}+{F\over \r} ({dy\over d\r})^2}}=C={\tt const}. \lb{**} \ee
The area of the minimal surface $\Gamma_h$  is then given by the
integral \be {\tt Area}(\Gamma_h)=2\pi
\int_{\epsilon^2}^{\r_m}d\r{\cal A} (\r),~{\cal A} ={R^2\over
\r^2\sqrt{1-{C^2\r^3\over FR^4}}} \lb{Ar} \ee Using (\ref{exp1})
we find that $ {\cal A} (\r)=[{r^2_+\over
\r^2}+{g_{(2)\theta\theta}(r_+)\over \r}+..]. $ Substituting this
expansion into (\ref{Ar}) we find that the first two terms produce
divergences (when $\epsilon$ goes to zero)  which, according to
our proposal, are to be interpreted as UV divergences of
entanglement entropy. At the black hole horizon, one has a
relation $2R_{\theta\theta}|_{r_+}=r_+^2(R-R_{aa})$. According to
the AdS/CFT dictionary, one has ${1\over G_N}={2N^2\over \pi}$,
where $N$ is number of colors in the boundary CFT. Putting
everything together and applying our proposal (\ref{proposal}), we
find for the divergent part (compare with general formula
(\ref{Sdiv1})) \be S_{\tt div}={A(\Sigma)\over 4\pi
\epsilon^2}N^2-{N^2\over 2\pi}\int_\Sigma({1\over 4}R_{aa}-{1\over
6}R)\ln\epsilon, \lb{Sdiv} \ee where $A(\Sigma)=4\pi r_+^2$ is the
horizon area.

Let's briefly discuss this result. The leading divergence in
(\ref{Sdiv}) has the form of the area law and resembles the
Bekenstein-Hawking entropy. This is  more than just a resemblance.
Replace the regularized boundary with a brane in a $Z_2$
configuration (expression (\ref{Sdiv}) should be multiplied by
factor of 2 then) as in \cite{Hawking:2000da}. The induced
Newton's constant on the brane is ${1\over G_4}={2N^2\over
\pi\epsilon^2}$. The first term in (\ref{Sdiv}) then (with factor
2 included) is exactly the Bekenstein-Hawking entropy $
S_{BH}={A(\Sigma)\over 4G_4} $ of a black hole on the brane. Thus,
our proposal supports the suggestion, originally made  in
\cite{Hawking:2000da} for de Sitter space-time, that the entropy
of a black hole
 on the brane is entirely  entanglement entropy.

The logarithmic term in (\ref{Sdiv}) is conformally invariant. It
is related to log divergence \cite{Henningson:1998gx}
$W_{log}={N^2\over 4\pi^2}\int[{1\over 8}R^2_{\mu\nu} -{1\over
24}R^2] \ln\epsilon$  in the quantum effective action of boundary
CFT. This relation is a particular manifestation of general
formula (\ref{Sdiv1}). In the case at hand one has
$c_1=-c_2/3=-{N^2\over 96\pi^2}$, $c_3=0$. In the free field limit
this set of values of the coefficients $c_k$ corresponds to ${\cal
N}=4$ $SU(N)$ super Yang-Mills multiplet.

The UV-finite terms in the entropy can be calculated in the limit
when the radial size $L_{\tt inv}=y(L)-y(0)$ of the subsystem
complimentary to the black hole is small. Then one can neglect the
dependence of the metric  on entire surface $\Gamma$ on  the
coordinate $y$. The area of the minimal surface is given by
 (\ref{Ar}) with
$ {\cal A}(\r )={R^2(r_+,\r)\over \rho^2\sqrt{1-({\r\over
\r_m})^3}}, $ where in $R^2(r_+,\r)=(g_{(0)}+g_{(2)}\r
+g_{(4)}\r^2+h_{(4)}\r^2\ln\r)_{\theta\theta}$ we drop terms of
order $\r^3$ and higher. The location of saddle point
$\r_m=(r_+^4/C^2)^{1/3}$ is determined from  the condition that
$\Gamma$ interpolates between spheres at $r=r_+$ and $r=L$,
 $L_{\tt inv}={1\over 2\r_m^{3/2}}\int_0^{\r_m}{d\r\r \over \sqrt{1-\r^3/\r_m^3}}$. Evaluation of the integral gives the
relation $\r_m={1\over \delta_1^2} L^2_{\tt inv}$,
$\delta_1={1\over 6}B({1\over 2},{2\over 3})$.

The $\r$-integration in (\ref{Ar}) can be performed explicitly and
the result is the sum of  UV-divergent terms (\ref{Sdiv}) and
finite terms. For the finite part we get an expansion in $L_{\tt
inv}$ (that generalizes (\ref{Lsm}) to a higher dimension), \be
S_{\tt fin}&=&{N^2\over 4\pi}(a_0L^{-2}_{\tt inv}+a_1\ln L_{\tt inv}+a_2 L^2_{\tt inv}\ln L_{\tt inv}) \lb{Sfin} \\
a_0&=&-\delta_1  {A(\Sigma)}~,~~a_1=\int_\Sigma({1\over 2}R_{aa}-{1\over 3}R), \nonumber \\
a_2&=&{\delta_2\over 48}\int_\Sigma[2(R-R_{aa})^2-\nabla^2
R-{3\over 2} Ric^2 +R^2_{aa}], \nonumber \ee where
$\delta_2={1\over 3 \delta_1^2}B({1\over 2},{1\over 3})$. The next
term in expansion (\ref{Sfin}) is $L^2_{\tt inv}$ with coefficient
determined by the value of $g_{(4)\theta\theta}$ on the horizon.
It is not expressed in terms of local functions of metric and
requires knowledge of the stress-energy tensor of boundary CFT.
The first two terms in (\ref{Sfin}) are local and repeat the
structure of the UV divergences (\ref{Sdiv}) so that, for a
certain value $L_{\tt inv}\sim \epsilon$, the total entropy
vanishes. The last term in (\ref{Sfin}) is of special interest.
This is the first term in the expansion (\ref{Sfin}) which
vanishes when $L_{\tt inv}$ goes to zero. The unusual presence of
the logarithm in this term should be noticed. Up to a numerical
factor, the coefficient $a_2$ is the value of the $(\theta\theta)$
component  of the ``obstruction tensor" (\ref{h4}) at the horizon.
(We have used relations $(\nabla_r
R_{\theta\theta})_{r_+}=r_+(2R_{aa}-R)_{r_+}$ and
$2(1-f'(r_+)r_+)=r^2_+(R-R_{aa})_{r_+}$ valid at the black hole
horizon when derived $a_2$ in the form (\ref{Sfin}).) As is well
known, the obstruction  tensor (despite its ultimate relation to
conformal anomaly) does not contribute to the quantum action since
it is traceless. Interestingly, it makes its direct appearance in
entanglement entropy.  This property takes place in any even
boundary dimension $d$ (this is the  dimension in which the term
$h_{({d})}\r^{d\over 2}\ln \r$ in expansion (\ref{exp1}) is
non-vanishing). The corresponding
 term in the entropy
takes the form (up to numerical factor) $S\sim
r_+^{d-2}h_{({d})\theta\theta}(r_+) L^2_{\tt inv}\ln L_{\tt inv}$.
The direct  calculation of such terms in entanglement entropy on
the CFT side is not yet available. The dual conformal field theory
in question is strongly coupled. For $d=4$ it is ${\cal N}=4$
super Yang-Mills theory and for $d=6$ it is a $(2,0)$ theory. Our
result (\ref{Sfin}) and its generalization for $d=6$ is, thus, a
prediction for the entropy in these theories.

I would like to thank V. Korepin for interesting discussions. This
work is supported in part by  DFG grant Schu 1250/3-1.

\end{document}